\documentstyle[prl,aps]{revtex}
\input BoxedEPS.tex
\SetepsfEPSFSpecial
\HideDisplacementBoxes

\begin{document}

\twocolumn[\hsize\textwidth\columnwidth\hsize
          \csname @twocolumnfalse\endcsname
\title{Carbon nanotubes: Ballistic transport or room-temperature 
superconductivity?} 
\author{Guo-meng Zhao$^{*}$} 
\address{ 
Department of Physics, Texas Center for Superconductivity and Advanced Materials, 
University of Houston, Houston, Texas 77204, USA}

\maketitle
\widetext
\vspace{0.3cm}

\begin{abstract}

We theoretically estimate the electron-phonon coupling constant $\lambda$ for 
metallic single-walled carbon nanotubes with a diameter of 1.4 nm. The partial 
electron-phonon coupling constant for the hardest phonon mode is 
estimated to be about 0.0036, in good agreement with that deduced from 
Raman scattering data assuming superconductivity above room 
temperature.  Assuming no superconductivity, we estimate the 
room-temperature inelastic mean free path $\l_{ep}$ due to 
electron-phonon scattering to be about 0.46~$\mu$m, and the total 
room-temperature inelastic mean free path $\l_{in}$ to be about 
0.16~$\mu$m.  We then argue that the electrical transport data of 
individual multi-walled nanotubes cannot be explained by ballistic 
transport at room temperature but provide strong evidence for 
quasi-one-dimensional superconductivity above room temperature.

~\\
\end{abstract}
\narrowtext

]

Our recent articles have provided strong evidence for 
very high-temperature superconductivity in both single-walled and 
multi-walled carbon nanotubes \cite{Zhao1,Zhao2,Zhao3,Zhao4}. The 
mean-field superconducting transition temperatures $T_{c0}$'s for 
both 
single-walled nanotubes (SWNTs) and multi-walled nanotubes (MWNTs) 
can be higher than 600 K. It has 
also been shown that the non-zero 
on-tube resistance 
state below $T_{c0}$ in some individual nanotubes is caused by 
quantum 
phase slips (QPS) inherent in quasi-one-dimensional (quasi-1D) 
superconductors 
\cite{Zhao3}, as observed in ultrathin wires of conventional 
superconductors such as PbIn and MoGe \cite{Giordano,Tinkham}.  The 
temperature dependence of the resistance in an individual SWNT or 
MWNT is very 
similar to that in the ultrathin wires of MoGe and can be naturally 
explained by the QPS theory \cite{Zhao3}.

Remarkably, negligible on-tube resistances have been observed 
in a multi-walled nanotube bundle consisting of two MWNTs with a 
diameter $d$ = 16 
nm (Ref.~\cite{Frank}), in an individual MWNT with $d$ = 40 nm 
(Ref.~\cite{Pablo}), in about 50 individual MWNTs \cite{Heer}, as 
well as in a single-walled nanotube bundle consisting of two SWNTs with 
$d$ = 1.4 nm (Ref.~\cite{Bachtold2000}). These electrical transport 
data have been tentatively explained 
in terms of ballistic transport at room temperature.

Here  we theoretically estimate the electron-phonon coupling constant $\lambda$ for 
metallic single-walled carbon nanotubes with a diameter of 1.4 nm. The partial 
electron-phonon coupling constant for the hardest phonon mode is 
estimated to be about 0.0036, in good agreement with that deduced from 
Raman scattering data assuming superconductivity above room 
temperature.  We also estimate the room-temperature inelastic mean 
free path $\l_{ep}$ due to electron-phonon 
scattering to be about 0.46~$\mu$m, and the total room-temperature inelastic 
mean free path $\l_{in}$ to be about 0.16~$\mu$m~assuming no superconductivity.  We then argue that 
the electrical transport data of individual multi-walled nanotubes 
cannot be explained by ballistic transport at room 
temperature but provide strong evidence for quasi-one-dimensional 
superconductivity above room temperature.

We first make a theoretical estimate of the electron-phonon coupling 
constant for carbon nanotubes.  Because there are some common 
features in 
graphites, C$_{60}$ and carbon nanotubes, one could estimate the 
electron-phonon coupling constant $\lambda$ for the carbon nanotubes 
from the known $\lambda$ values of graphites and C$_{60}$.  It was 
argued 
that \cite{Bend} curvature-induced hybridization in both C$_{60}$ and 
carbon 
nanotubes opens additional electron-phonon scattering channels that 
are not available to flat graphite sheets.  Neglecting second-order effects, the electron-phonon pairing potential can be 
decomposed into two components, one of which is present in flat sheet, 
and the other arising from new scattering channels due to non-zero 
curvature.  
The pairing potential in C$_{60}$ is \cite{Bend}
\begin{equation}
U_{ball} = U_{flat} + U_{curve}(\frac{R_{\circ}}{R})^{2},
\end{equation}
where $U_{curve}$ is the curvature contribution to the pairing 
potential for a ball of radius of $R_{\circ}$. Since the contribution 
to the electron-phonon matrix element from new scattering channels 
in single-walled nanotubes is half the size of the contribution in 
C$_{60}$, one readily shows that \cite{Bend}
\begin{equation}
U_{tube} = U_{flat} + \frac{1}{4}U_{curve}(\frac{R_{\circ}}{R})^{2}.
\end{equation}

For doped C$_{60}$, four independent calculations lead to a similar 
electron-phonon 
coupling constant \cite{Gunnarsson}. The average electron-phonon 
coupling constant 
$\lambda = N(0)U$ from the four independent calculations is $\lambda$ 
= 0.40 assuming that the density of 
states at the Fermi level $N(0)$ = 0.24 states per carbon atom per eV 
(Ref.~\cite{Gunnarsson}). 
The contribution from the hardest mode ($\hbar\omega$ = 197 meV) is 
$\lambda^{h}$ = 0.09 (Ref.~\cite{Gunnarsson}).  Further, the coupling 
constant for doped 
graphites with $N(0)$ = 0.24 states per carbon atom per eV is 0.25 
(Ref.~\cite{Bend}). 
Substituting the above values into Eq.~1 yields $U_{curve}$ = 
0.6$U_{flat}$. For SWNTs with $R$ = 0.7 nm, $N(0)$ = 0.015 states per 
carbon atom per eV (Ref.~\cite{Mintmire}). The low density of states 
in the SWNTs implies a 
low electron-phonon coupling constant.  Using Eqs.~1-2 and the 
relation $R$ $\simeq$ 2$R_{\circ}$, we can easily show 
that the electron-phonon coupling constant for the 1.4~nm~diameter 
SWNTs is a 
factor of $f$ = 24.7 smaller than that for doped C$_{60}$.  With this 
reduction factor, we finally find that the total electron-phonon 
coupling constant for the 1.4~nm~diameter SWNTs is $\lambda$ = 0.016, 
and the contribution to the electron-phonon coupling constant from the hardest phonon mode 
$\lambda^{h}$ = 0.0036.

The electron-phonon coupling constant for a particular phonon mode 
could be  estimated from the phonon self-energy effects due to 
superconducting pairing. The phonon self-energy effects due to 
superconductivity can lead to shifts in both 
the frequency and width of the phonon modes if the phonons are coupled 
to conduction electrons.  Such 
phonon self-energy effects were 
clearly demonstrated in the high-$T_{c}$ cuprate superconductors 
\cite{Krantz,Cardona,Ham}. The width shifts of the Raman 
active phonon modes in  90 K 
superconductors RBa$_{2}$Cu$_{3}$O$_{7-y}$ (R is a rare-earth 
element) 
are in quantitative agreement with the theoretical 
calculation \cite{Cardona}. 

\begin{figure}[htb]
\input{epsf}
\epsfxsize 7cm
\centerline{\epsfbox{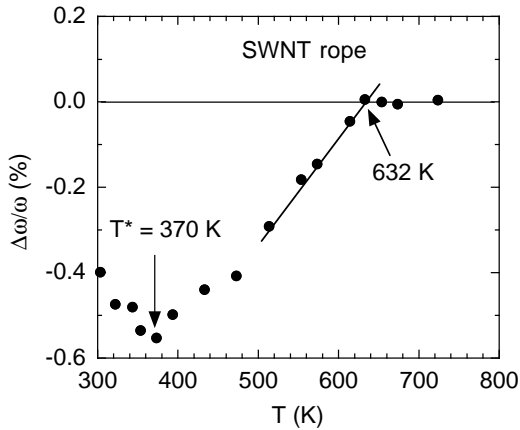}}
	\vspace{0.6cm}
	\caption [~]{The temperature 
dependence of the relative frequency shift for the hardest Raman 
active mode with $\hbar\omega$ = 197 meV. The data are from R.  
Walter {\em et al.} at the University of North Carolina 
\cite{Walter}.  
The detailed procedure for obtaining $\Delta\omega$  has been 
described in 
Ref.~\cite{Zhao4}.}
	\protect\label{fig1}
\end{figure}

The phonon self-energy effects were also 
seen in the hardest Raman-active $G$-band ($\hbar\omega$ = 197 meV) 
of single-walled carbon nanotubes with superconductivity above 600 K 
(Ref.~\cite{Zhao4}).  In Fig.~1, we plot the temperature 
dependence of the relative frequency shift for the hardest Raman 
active mode with $\hbar\omega$ = 197 meV.  The data are from R.  
Walter {\em et al.} at the University of North Carolina 
\cite{Walter}.  
The detailed procedure for obtaining $\Delta\omega$  has been 
described in 
Ref.~\cite{Zhao4}.  It is apparent that the frequency starts to shift 
down at 632 K, reaching to a minimum at $T^{*}$ = 370 K.  Such a 
temperature dependence is similar to the calculated one where the 
frequency starts to decrease below 0.95$T_{c0}$ and reaches a 
shallow minimum at about 0.6$T_{c0}$ (see Fig.~8 of 
Ref.~\cite{ZZ1} in the case of $\hbar\omega$/$2\Delta (0)$ = 0.88, 
where $\Delta (0)$ is the gap at zero temperature).  The shallow 
minimum in the frequency shift is related to a sharp minimum in the 
real part of the polarization $\Pi(\omega,T)$, which occurs at 
$\hbar\omega$/$2\Delta (T^{*})$ = 1.05 for strong coupling and at 
$\hbar\omega$/$2\Delta (T^{*})$ = 1.0 
for weak coupling \cite{ZZ1,ZZ2}. The weaker the coupling, the sharper 
the minimum in the real part of the polarization \cite{ZZ2}, and thus 
the 
more pronounced the minimum in the frequency shift.  If we assume 
weak coupling and take $T_{c0}$ 
= 660 K (because 0.95$T_{c0}$ = 630 K), we obtain $T^{*}$ = 
0.57$T_{c0}$, and thus $\Delta (T^{*})$ = 
0.93$\Delta (0)$ from the BCS theory.  Then $2\Delta (0)$ = 1.07$[2\Delta (T^{*})]$ = 
1.07$\hbar\omega$ = 210 meV, in excellent agreement with that deduced 
from the independent tunneling spectrum \cite{Zhao3}.

We can evaluate the contribution to the electron-phonon coupling constant 
from the hardest phonon mode $\lambda^{h}$ using a 
relation \cite{Cardona}: $\Delta\omega/\omega$ = 
0.5$\lambda^{h}$Re$\Pi/N(0)$.  The theoretical 
calculations \cite{ZZ1,ZZ2} suggest that Re$\Pi/N(0) \simeq -2$ at 
$T^{*}$ = 
0.5$T_{c0}$ for strong coupling, and that Re$\Pi/N(0)$ becomes more 
negative when the coupling decreases.  Therefore, we may expect that 
the value of 
Re$\Pi/N(0)$ should be in the range between  $-2$ and $-$3 at $T^{*}$ = 
0.56$T_{c0}$ for weak coupling.  Then we obtain $\lambda^{h}$ = 
0.0037-0.0056, which is consistent with the theoretical estimate above. 
This consistency suggests that the unusual temperature dependence of 
the frequency for the hardest phonon mode of the single-walled nanotubes 
is indeed related to superconductivity above 600 K, and that the 
theoretically estimated coupling constant $\lambda$ is reliable.

From the values of the total coupling constant $\lambda$ and the 
Debye temperature $\theta_{D}$, we can now calculate the inelastic 
electron-phonon scattering time $\tau_{ep}$ using an equation 
\cite{Allen}
 \begin{equation}\label{EP}
\frac{1}{\tau_{ep}} = 
24\pi\xi(3)\frac{\lambda}{1+\lambda}\frac{k_{B}}{\hbar}\frac{T_{e}^{3}}{\theta_{D}^{2}},
\end{equation}
where $\xi (3)$ = 1.2 and $T_{e}$ is the temperature of electrons. Usually, the value of $\theta_{D}$ is about 
1.2$\hbar\omega_{\ln}/k_{B}$ (Ref.\cite{ZhaoNP}).  With 
$\hbar\omega_{\ln}/k_{B}$ =1400 K (Ref.~\cite{Bend}), 
we have $\theta_{D}$ = 1680 K.  
Substituting $T_{e}$ = 300 K, $\lambda$ = 0.016, and $\theta_{D}$ = 1680 
K into Eq.~\ref{EP}, we obtain $\tau_{ep}$ = 0.56 ps at room temperature.

On the other hand, the room-temperature electron-phonon scattering 
time $\tau_{ep}$ = 18 ps was estimated from the measured time 
dependence of 
electron temperature $T_{e}(t)$ (Ref.~\cite{HertelPRL}).  This 
value is larger than the above theoretical estimate by  a factor of 
about 30.  The recent work by the same group \cite{Hertel} appears to indicate 
that $\tau_{ep}$ is much smaller than their previous estimate and that 
the value of $\tau_{ep}$ depends strongly on the data sets.  In fact, 
the ways to determine $\tau_{ep}$ in Ref.~\cite{HertelPRL} and 
Ref.~\cite {Hertel} are not robust.  In Ref.~\cite {Hertel}, the 
authors fit their high-temperature data ($>$ 400 K) with a formula 
that is only valid at low temperatures ($T_{e} << \theta_{D}$).  
Accordingly, the value of $\tau_{ep}$ deduced from the fit to the 
high-temperature data is obviously unreliable.  On the other hand, the 
data analysis in Ref.~\cite{HertelPRL} may be justified if there were 
no superconductivity in those nanotubes.  The basic idea in the data 
analysis of Ref.~\cite{HertelPRL} is that, if there is no superconductivity, $\tau_{ep}$ is 
inversely proportional to $q_{0}^{2}$ and $q_{0}$ is related to 
$T_{e}(t)$ by $dT_{e}/dt$ $\propto$ $[N(0)q_{0}]^{2}/C_{e}$ 
(Ref.~\cite{HertelPRL}), where $C_{e}$ is the normal-state electronic 
specific heat and $q_{0}$ is the relative change of the 
nearest-neighbor hopping integral $\gamma_{\circ}$ with respect to 
lattice distortion.  Assuming $N(0)$ is equal to a theoretically 
predicted value, Hertel and Moos \cite{HertelPRL} deduced $q_{0}$ = 
0.8 \AA$^{-1}$ from the data of $T_{e}(t)$ in the temperature range of 
500-700 K.  This deduced value of $q_{0}$ is about three times smaller 
than the calculated one (2.5 \AA$^{-1}$) that was confirmed by an 
independent experiment on a graphite-related compound \cite{Pie}.  
There are no reasons to believe that $q_{0}$ for carbon nanotubes 
should be smaller than for other graphite-related materials since this 
quantity is determined by the local chemical bonding that should 
remain unchanged in graphite-related materials.

This apparent discrepancy can be resolved if one assumes that the SWNT 
rope may exhibit superconductivity above 600 K, as demonstrated from 
many other independent experiments \cite{Zhao3,Zhao4}.  In the 
superconducting state, one may expect that 
$dT_{e}/dt$ $\propto$ $[N_{s}q_{0}]^{2}/C_{es}$, where $N_{s}$ 
and $C_{es}$ are the  quasi-particle density of states and the electronic 
specific heat below $T_{c0}$, respectively. According to the 
BCS theory, $N_{s}^{2}/C_{es}$ is much smaller 
than its counterpart $[N(0)]^{2}/C_{e}$ in the normal state, and goes to zero at zero temperature. This is because for 
0.5$T_{c0}$$<$$T$$<$0.9$T_{c0}$, $N_{s}$ is much smaller than 
$N(0)$ 
and $C_{es}$ $>$ $C_{e}$, while for $T$$<$ 0.2$T_{c0}$ both 
$N_{s}(0)$ 
and $C_{es}$ tend to zero so that $N_{s}^{2}/C_{es}$ tends to zero 
\cite{Book}. Just slightly below $T_{c0}$, $N_{s}$$\simeq$ $N(0)$ 
and $C_{es}$ = 2.43$C_{e}$ such that $N_{s}^{2}/C_{es}$$\simeq$ 
(1/2.43)$[N(0)]^{2}/C_{e}$. This scenario can 
naturally explain why $dT_{e}/dt$ tends to zero even when $T_{e}$ is 
higher than the lattice temperature by more than 100 K (see Fig.~3 of 
Ref.~\cite{HertelPRL}). In fact, the initial slope 
$dT_{e}/dt$ at $T_{e}$$\simeq$ 730 K (see Fig.~3 
of Ref.~\cite{HertelPRL}) is consistent with $T_{c0}$ $\simeq$ 730 K and 
$q_{0}$ $\simeq$ 2 \AA$^{-1}$.  If we use $T_{c0}$ $\simeq$ 730 K and 
$q_{0}$ $\simeq$ 2 \AA$^{-1}$, we 
find that the data in the whole temperature range can be well fitted by 
$N_{s} (T) = 2N(0)/(\exp [\Delta (T)/k_{B}T] +1)$.

By taking $q_{0}$ = 2.5 \AA$^{-1}$, $\gamma_{\circ}$ = 2.5 eV, and 
using Eq.~28 of Ref.~\cite{Jishi}, we find that the room-temperature 
electron-phonon scattering time due to the longitudinal acoustic 
phonon is 1.05 ps.  
Considering the contributions of the electron-phonon scattering from 
other phonon modes, one expects that the total $\tau_{ep}$ $<$ 
1.05 ps, in agreement with the independent theoretical estimate above.

With $v_{F}$ = 
8$\times$10$^{5}$m/s (Ref.~\cite{Mintmire}), the room temperature 
inelastic mean free path 
due to electron-phonon scattering is $\l_{ep}$ = 0.46 $\mu$m.  Since 
$\l_{ep}$ is inversely proportional to the density of states per 
carbon which in turn  is itself inversely proportional to the tube diameter, 
$\l_{ep}$ is proportional to $d$,
\begin{equation}\label{L}
\l_{ep}= 0.46 \frac{d}{d_{\circ}}~(\mu m).
\end{equation}
where $d_{\circ}$ = 1.4 nm.

We now discuss the electrical transport of carbon nanotubes. The 
scattering properties of the metallic subbands (the central two 
subbands of metallic tubes) and the semiconducting subbands (the 
bands 
that do not cross the Fermi level of undoped metallic tubes) are very 
different.  It was shown that the electron backscattering from a single 
impurity with long range potential is nearly absent in metallic 
subbands while this backscattering becomes significant for 
semiconducting subbands \cite{RocheAPL}.  When the Fermi level 
crosses 
the semiconducting subbands in doped metallic SWNTs, one should 
expect 
that the two metallic subbands provide the main contribution to the 
electrical transport.  Indeed, the elastic mean free path $\l_{el}$ 
for individual metallic SWNTs with $d$ = 1.4 nm was theoretically 
estimated to be 
about 3 $\mu$m, implying a possibility of ballistic transport 
\cite{White}.

On the other hand, the situation in multi-walled nanotubes becomes much 
more complicated. The chiralities of different shells constituting a 
MWNT play an important role in electrical transport. If several 
shells 
have the same chiral angle, their periodicities along the nanotube 
axis 
are commensurate. If chiral angles of different shells are varied, 
their 
periodicities are incommensurate, which should be case in real 
systems \cite{Roche}.   In the absence of defects, the intershell 
coupling allows 
the electron to spread over several shells, and when
their periodicities  are incommensurate, the
electrical transport is theoretically shown to be non-ballistic 
\cite{Roche}. Only 
if all the shells are commensurate, which is very unlikely in real 
materials,  can ballistic transport be a possibility \cite{Roche}.  However, 
defects and 
disorders are inevitable in real materials. The electron 
backscattering 
is shown to be much more 
pronounced in a MWNT which consists of commensurate metallic chirality shells 
than in a MWNT with alternating metallic 
and semiconducting chirality shells \cite{RocheAPL}.  Therefore, electrical 
transport in MWNTs 
should not be ballistic unless there are no defects and at least two 
adjacent shells have the same metallic chiral angle.

Although the probability of producing a defect free MWNT which has 
more than two adjacent shells with the same  metallic chiral angle 
is low, we could calculate the resistance 
per unit length at room temperature for a model MWNT where the number 
of outermost adjacent layers with the same  metallic chiral angle is $N_{m}$.  
At most, $N_{m}$ outermost shells have an infinitely long elastic 
mean-free path 
so that the total mean free path in each of these shells is equal to 
the inelastic mean free 
path. Considering that $N_{m}$ shells mainly contribute to electrical 
transport due to the fact that the transport in other incommensurate 
shells is non-ballistic, the 
resistance per unit length of the model MWNT at room 
temperature is then given by \cite{Bend}
\begin{equation}\label{LE}
\frac{R}{L}= \frac{R_{Q}}{2N_{m}\l_{in}},
\end{equation}
where $R_{Q} = h/2e^{2}$ = 12.9 k$\Omega$, Since 
there are inelastic electron-electron, electron-plasmon, and 
electron-phonon scattering, we expect that $\l_{in}$ $<$ $\l_{ep}$.  
Using Eq.~\ref{L} and the above inequality, we find 
\begin{equation}\label{R}
\frac{R}{L} > \frac{14000d_{\circ}}{N_{m}d}~(\Omega\mu m^{-1}). 
\end{equation}

In Fig.~2, we plot the calculated lower limits of the room-temperature $R/L$ as a function 
of $d$ for $N_{m}$ = 1, 3, and 10 in the case of nonsuperconducting 
transport. The upper limits of the measured 
$R/L$ are also included in Fig.~2 for comparison. These two data are 
taken from Refs.~\cite{Pablo,Heer}.  In Ref.~\cite{Heer}, extensive 
transport investigations have been carried out on MWNTs with 
diameters ranging from 5 to 25 nm. These MWNTs which protrude 
from unprocessed arc produced nanotube fibers and are contacted with 
liquid metals show small per unit length resistances $R/L$ $<$ 100 
$\Omega$/$\mu$m at room temperature.  For example,  $R/L$ 
=14 $\Omega$/$\mu$m for a MWNT with $d$ $\simeq$ 20 nm.  In 
Ref.~\cite{Pablo}, a reliable technique of 
making ideal Ohmic contacts on MWNTs was developed to 
study electrical transport in individual MWNTs. It was 
found that the on-tube room temperature resistance per unit length 
$R/L$ 
$<$0.5 $\Omega$/$\mu$m for a MWNT with $d$ = 40 nm. 
\begin{figure}[htb]
\input{epsf}
\epsfxsize 7cm
\centerline{\epsfbox{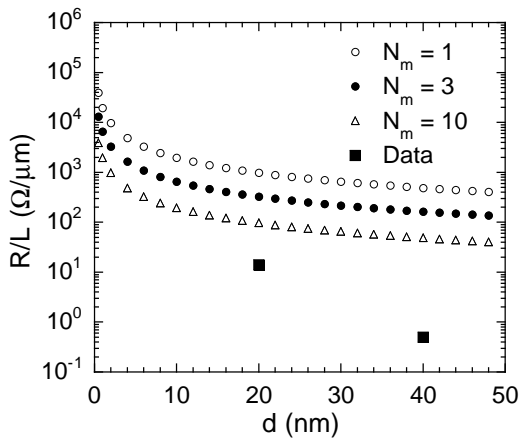}}
	\vspace{0.6cm}
	\caption [~]{The calculated lower limits of the room-temperature $R/L$ as a function 
of $d$ for $N_{m}$ = 1, 3, and 10, in the case of nonsuperconducting 
transport. The upper limits of the measured 
$R/L$ are also included for comparison. These two data  are 
taken from Refs.~\cite{Pablo,Heer}.}
	\protect\label{fig2}
\end{figure}

From Fig.~2, one 
can clearly see that the measured resistance per unit length at room 
temperature is even far below the calculated lower limit for 
$N_{m}$ = 10. Since the probability of producing a defect free MWNT 
which has 
10 outermost shells with the same metallic chiral angle is nearly 
zero, the electrical transport data reported in 
Refs.~\cite{Pablo,Heer} 
cannot be explained if MWNTs were not room temperature 
superconductors. 
Since the periodicities for the majority of MWNTs should be 
incommensurate 
and non-ballistic in real world \cite{Roche}, it is very difficult to 
understand how such small per unit length 
resistances ($R/L$ $<$ 100 
$\Omega$/$\mu$m) at room temperature in the majority of the MWNTs 
studied \cite{Heer} could arise if the MWNTs were {\em not} room temperature 
superconductors.
\begin{figure}[htb]
\input{epsf}
\epsfxsize 7cm
\centerline{\epsfbox{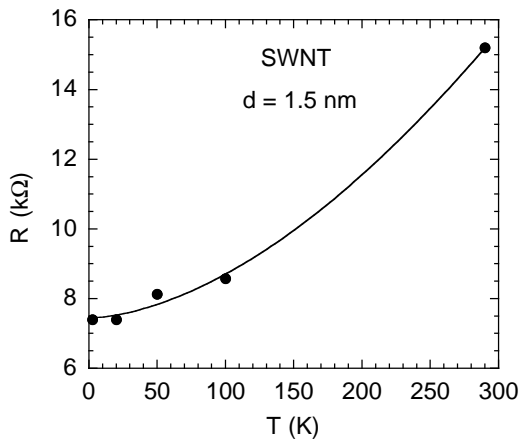}}
	\vspace{0.6cm}
\caption [~]{Temperature dependence of the resistance for a 
single-walled nanotube with $d$ = 1.5 nm.  The data are extracted from 
Fig.~1a of Ref.~\cite{Kong} at zero gate voltage.}
	\protect\label{fig3}
\end{figure}

The calculated curves in Fig.~2 are the lower limits of $R/L$ since 
we have only considered the contribution from electron-phonon 
scattering, which is very weak and cannot lead to high-temperature 
superconductivity. In fact, electron-plasmon coupling in quasi-1D 
systems is strong and could result in superconductivity as high as 
500 
K \cite{Lee}. We can estimate the total room-temperature inelastic 
mean free 
path  from electrical transport data assuming no superconductivity. Fig.~3 shows the temperature 
dependence of the resistance for a 
single-walled nanotube with $d$ = 1.5 nm. The data are extracted from 
Ref.\cite{Kong}. The distance between the two contacts is about 200 
nm 
and the contacts are nearly ideal with the transmission probability 
of 
about 1 \cite{Kong}. From Fig.~3, we can see that the resistance 
increases with increasing temperature. The temperature dependent part 
of the resistance should arise from the inelastic scattering if 
there were no superconductivity. Then we find that the 
room-temperature resistance per unit length due to 
inelastic scattering is about 40 k$\Omega$/$\mu$m (the difference 
between the resistance at room-temperature and the one at zero 
temperature). Substituting this value into Eq.~\ref{LE} yields 
$\l_{in}$ = 0.16 $\mu$m at room temperature. 
With such a short inelastic mean free path at room temperature and 
using Eq.~5, one finds that the on-tube resistance at room 
temperature across a length of 1 $\mu$m would be larger than 20 k$\Omega$ for 
a SWNT bundle that consists of two coupled SWNTs \cite{Bachtold2000}.  
This is in sharp contrast with a negligible on-tube resistance ($<$3 
k$\Omega$) observed in this bundle \cite{Bachtold2000}. The more plausible explanation is that the 
metal-like temperature dependence of the resistance shown in Fig.~3 is 
not caused by inelastic scattering but is associated with the 
quantum-phase slips, as discussed in Ref.\cite{Zhao3,Zhao4}.  The 
Josephson coupling between two SWNTs suppresses quantum phase slips 
and reduces the resistance of the bundle to a value that is remarkably 
smaller than the sum of the individual resistances.

Quantum-phase slip theory \cite{Zaikin} can also naturally 
explain 
the semiconductor-like behavior below room temperature in individual 
MWNTs that are lithographically contacted \cite{Sheo}.  The 
lithographically 
contacted MWNTs may contain high density of defects or imperfections 
which are introduced through purification and other processing steps 
\cite{Heer}. Defects and disorders enhance quantum-phase slips and 
the localization of Cooper pairs, which would lead to a 
semiconductor-like 
temperature dependence of the resistance below $T_{c0}$ 
(Ref.~\cite{Zaikin,Zhao3}).

In summary,  we theoretically estimate the electron-phonon coupling constant $\lambda$ for 
metallic single-walled carbon nanotubes with a diameter of 1.4 nm. The partial 
electron-phonon coupling constant for the hardest phonon mode is 
estimated to be about 0.0036, in good agreement with that deduced from 
Raman scattering data assuming superconductivity of about 700 K. Assuming 
no superconductivity, we estimate the room-temperature inelastic mean 
free path $\l_{ep}$ due to electron-phonon scattering to be about 
0.46~$\mu$m, and the total room-temperature inelastic mean free path 
$\l_{in}$ to be about 0.16~$\mu$m.  We further demonstrate that the 
electrical transport data of individual MWNTs cannot be explained by 
ballistic transport at room temperature but instead provide strong 
evidence for quasi-1D superconductivity above room temperature.

 ~\\
{\bf Acknowledgment:} I am grateful to Dr. R. Walter {\em et al.} for 
sending me 
their unpublished data, and to Prof. W. A. de Heer for sending me their preprint 
\cite{Heer}. I thank Dr. Pieder Beeli for bringing my 
attention to Ref.~\cite{Heer}, and for his critical reading and 
comments on the manuscript. The author acknowledges financial support from the State of Texas 
through the Texas Center for Superconductivity and Advanced Materials at 
the University of Houston where some of the work was completed. 
 ~\\
 ~\\  
* Correspondence should be addressed to gmzhao@uh.edu.

\end{document}